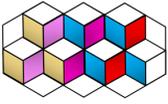
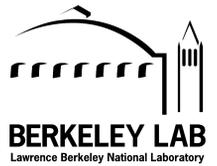

# Detecting Partial Rosettes in Tumor Histopathology Using The Cross Product

David H. Nguyen, Ph.D.
Principal Investigator
Tissue Spatial Geometrics Laboratory

Affiliate Scientist
Department of Cellular & Tissue Imaging
Division of Molecular Biophysics & Integrated Bioimaging
Lawrence Berkeley National Laboratory

DHNguyen@lbl.gov

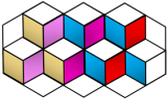

## Abstract

Tumors of the eye and nervous system often exhibit an arch-like arrangement of nuclei, called rosettes. Pathologists are able to identify rosettes (full circles) and the presence of partial rosettes (semi-circles) and interpret this as a sign of differentiation in a tumor. However, there is no objective method to quantitate the many partial rosettes that are obvious or not obvious to the naked eye. This paper proposes a mathematical algorithm to computationally detect the presence of obvious or non-obvious partial rosettes, henceforth referred to as "Nguyen-Wu Partial Rosettes." Quantifying the degree of partial rosettes present in a tumor may allow pathologists to stratify tumors into more refined groups that may respond better to therapy or have different clinical outcomes. The Midline Cross Product (MCP) algorithm calculates the magnitude of two cross products and adds them together to obtain one value. Each of the two cross products results from (1) the line that connects the midpoints of longest lengths of two neighboring ovals, and (2) the line that extends from the midpoint from one longest length and is perpendicular to that longest length. The MCP algorithm makes nuclei that are arranged in consecutive rows and arches quantitatively distinct from nuclei that are arranged next to each other in a disorderly manner.

## Graphical Abstract

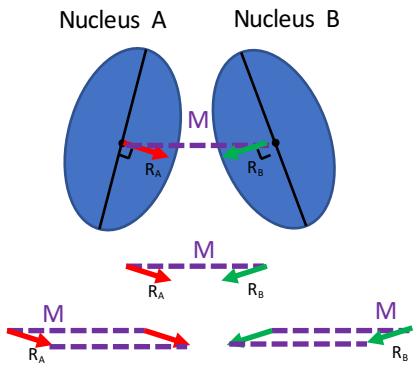

Midline Cross Product (MCP) = ($R_A$ X M) + ($R_B$ X M)

M = length of line between the midpoints of the longest lengths of nucleus A and nucleus B

$R_A$ and $R_B$ = (M/2) = a vector originating at the midpoint of a nucleus; whose magnitude is ½ the length of M; and whose direction is orthogonal to the longest length of the nucleus

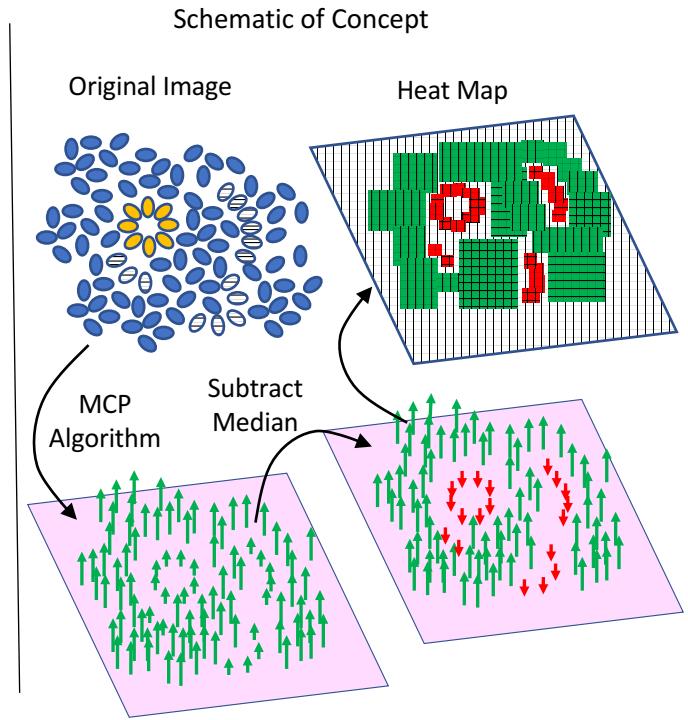

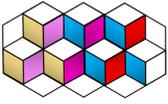

**Background**

Tumors of the eye and nervous system often exhibit a pattern of nuclei arrangement that is known as a rosette (Figure 1). Full rosettes are obvious to the naked eye, so pathologists are able to score the level of differentiation of a tumor based on their presence (Figure 2). However, these tumors can also exhibit partial rosettes that are not obvious to the naked eye. Even if partial rosettes are visually identifiable, their characteristics and abundance are too complex to be assessed by a pathologist. However, a computational approach would be able to quantify the degree to which partial rosettes are present, and the distribution of how many short, medium, or long partial rosettes are present.

This paper proposes that partial rosettes can be routinely identified by the algorithm defined here as the Midline Cross Product. This method has the potential to stratify tumors that contain partial rosettes into refined categories that may respond better to chemotherapy or that warrant specific types of chemotherapy. This refined information will spare patients of unnecessary treatments, which can have severe negative side effects. These partial rosettes are referred to as Nguyen-Wu Partial Rosettes.

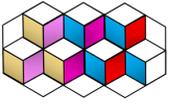

# Figure 1

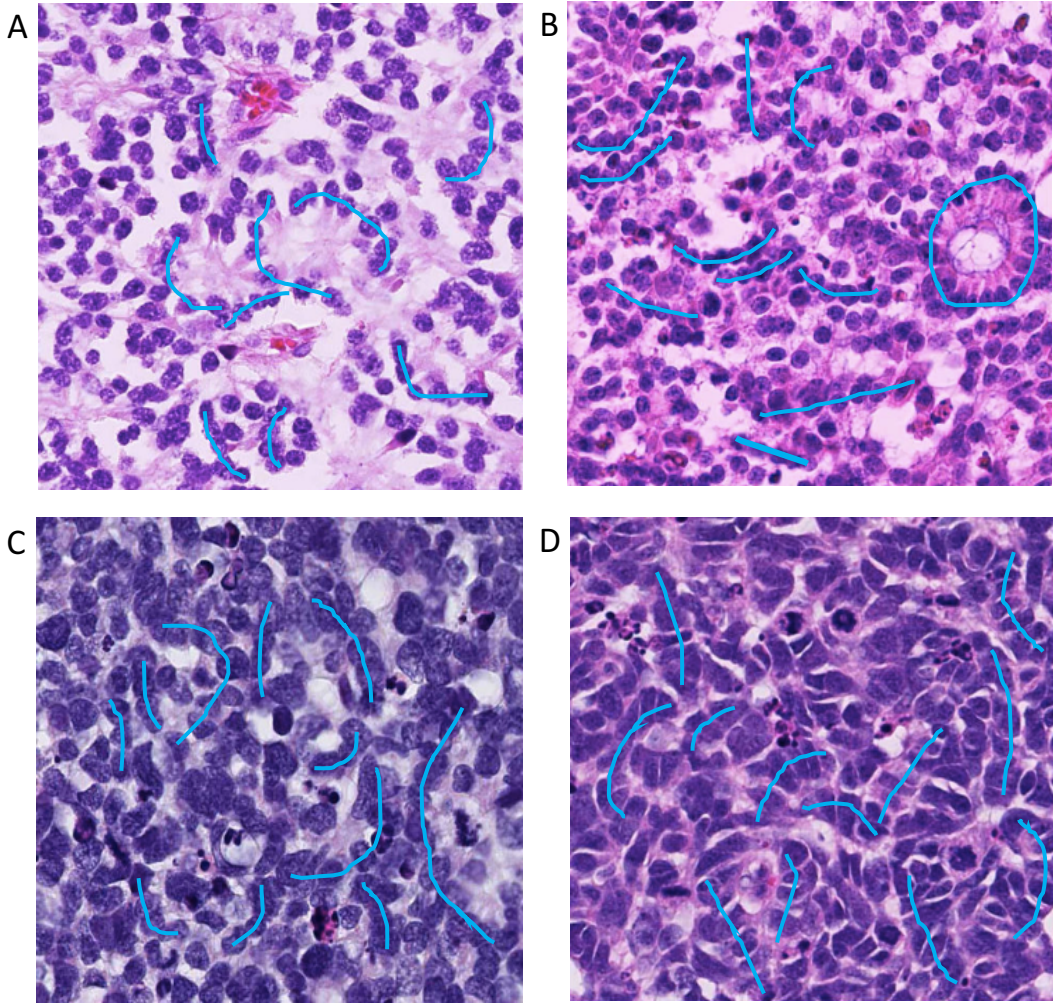

**Figure 1. Rosettes in Eye Tumors.** Blue lines trace examples of partial rosettes in each image. A full rosette is the blue circle in B. (A) Retinocytoma. (B) Retinoblastoma with mild anaplasia. (C) Retinoblastoma with moderate anaplasia. (D) Retinoblastoma with severe anaplasia. Images adapted from Mendoza et al. 2015.

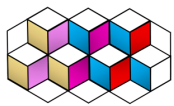

# Figure 2

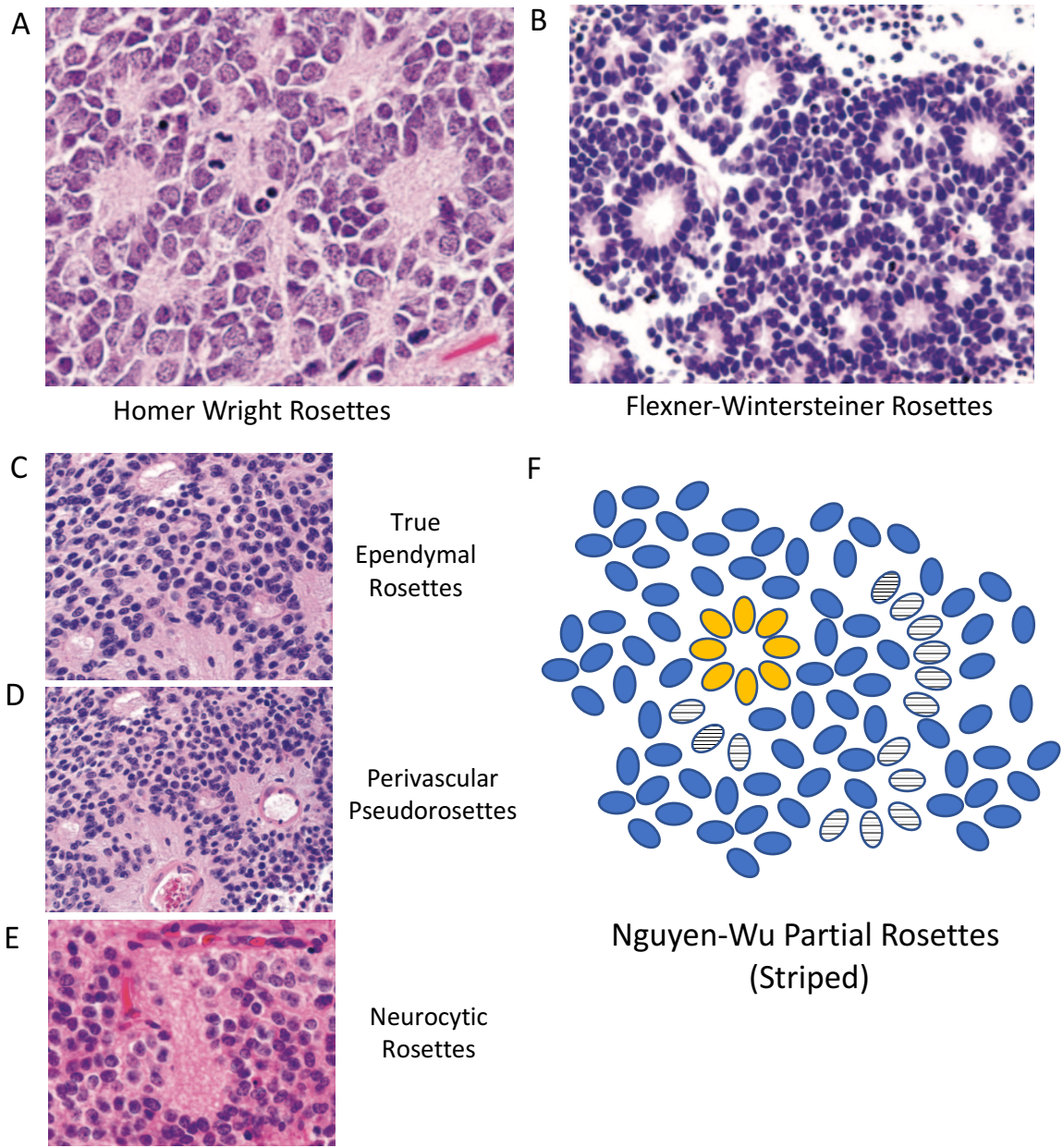

**Figure 2. Types of Rosettes in Tumor Histopathology.** (A) Homer Wright rosettes common in medulloblastoma. (B) Flexner-Wintersteiner Rosettes common in retinoblastoma and pineoblastoma. (C) True ependymal rosettes in ependymoma. (D) Perivascular pseudorosettes found in ependymoma. (E) Neurocytic rosettes found in neurocytoma. (F) This paper proposes the MCP algorithm for detecting partial rosettes via computational methods, and calls them Nguyen-Wu Partial Rosettes. A-E were adopted from Wippold & Perry 2006.

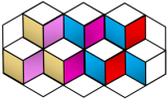

## Methodology

Neighboring nuclei that form a straight row or an arched row can be identified as distinct quantitative features compared to the nuclei that are arranged next to each other in a disorderly fashion. This paper presents an adaptation of the cross product of two vectors for detecting adjacent nuclei that form rows or arches. The algorithm will be referred to as the Midline Cross Product (MCP) algorithm and is explained in detail in Figure 3 and its legend. In summary, adjacent nuclei that are arranged in a row or an arch will have a small MCP value (Figure 4) compared to adjacent nuclei that are not in a row or arch arrangement (Figure 5).

The cross product can be calculated geometrically (as explained below), or algebraically through vector multiplication.

Geometrically: The magnitude of a cross product is equivalent to the area of the parallelogram whose sides are the two vectors of the cross product. The formula for the area of a parallelogram is:

[Equation 1] area = base x height

The length of the base is one of the two vectors of the cross product. The height is unknown, but can be calculated if the acute angle (θ) between the two vectors is known. The height can be calculated by the formula:

[Equation 2] height = (length of slant) x sin(θ)

Since the length of the slant is equivalent to the length of the vector that is not the base of the parallelogram:

[Equation 3] height = (length of the vector that is not the base) x sin(θ)

Knowing the height of the parallelogram from Equation 3 allows us to calculate the area of the parallelogram as in Equation 1.

The method in this paper ignores the right-hand rule that determines the direction of the result of a cross product. By taking the absolute value of any cross product that is calculated, the MCP algorithm ensures that all MCP values are positive and point in the same direction. This allows for small MCP values in a dataset to become close to zero or negative after the dataset is median centered (Figure 6).

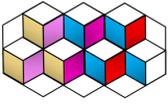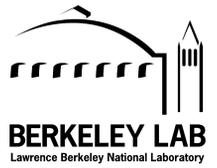

**Acknowledgements**
The author would like to thank Dr. Yarr Wu, O.D., for discussions about eye structure and diseases. This work would not have happened without inspiration from Dr. Wu to apply geometric algorithms to ocular histopathology.

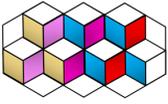

Tissue Spatial Geometrics Laboratory

BERKELEY LAB
Lawrence Berkeley National Laboratory

# Figure 3

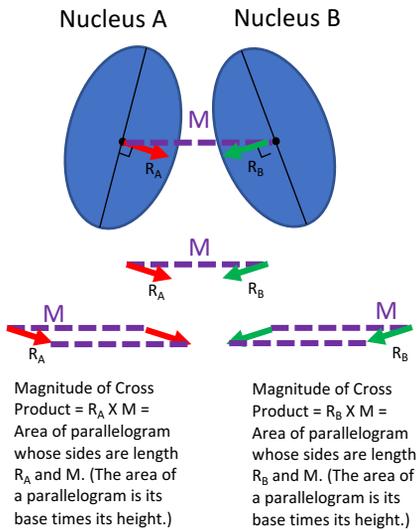

Nucleus A   Nucleus B

Magnitude of Cross Product = $R_A \times M$ = Area of parallelogram whose sides are length $R_A$ and M. (The area of a parallelogram is its base times its height.)

Magnitude of Cross Product = $R_B \times M$ = Area of parallelogram whose sides are length $R_B$ and M. (The area of a parallelogram is its base times its height.)

Midline Cross Product (MCP) = $|(R_A \times M)| + |(R_B \times M)|$

M = length of line between the midpoints of the longest lengths of nucleus A and nucleus B

$R_A$ and $R_B$ = (M/2) = a vector originating at the midpoint of a nucleus; whose magnitude is ½ the length of M; and whose direction is perpendicular to the longest length of the nucleus

Under the parameters defined as M and $R_{A/B}$, two adjacent ellipsoid nuclei that align parallel to or nearly parallel to each other will have an MCP whose magnitude is smaller than the MCP of two nuclei that have a different adjacent arrangement. Taking the absolute value of each cross product bypasses the right-hand rule and ensures that all MCP values are positive.

**Figure 3. Defining the Midline Cross Product (MCP).** The MCP algorithm was formulated to capture the arch-like arrangement of nuclei in rosettes. At least three consecutive adjacent nuclei are required to form an arch. (A) The midline "M" is the line connecting the midpoints of two lines. The lines are the longest lengths of each adjacent oval. The vectors $R_A$ (red) and $R_B$ (green) are perpendicular to their respective longest length lines. For consistency, the magnitude of $R_A$ and $R_B$ are defined to by half of M, so that the factors affecting the value of an MCP calculation are just the length of M and the angles between M and $R_A$, or M and $R_B$. If warranted, the values of $R_A$ and $R_B$ can be modified to be equal to half that of the longest lengths of their respective nuclei. This modification will allow the longest length of each nuclei to affect the MCP value.

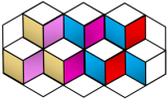
Tissue Spatial Geometrics Laboratory

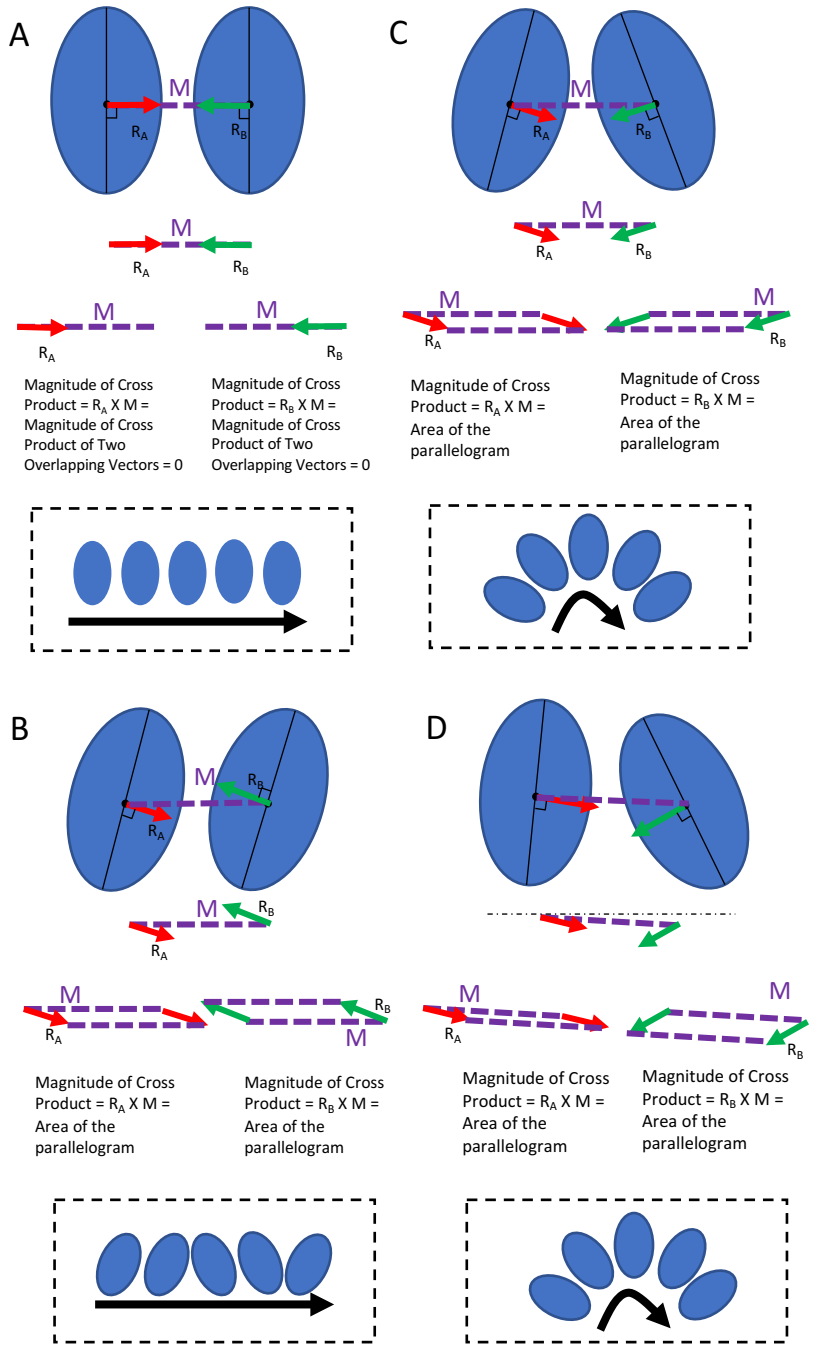

Figure 4

**Figure 4. Special Cases of Neighboring Nuclear Arrangements.**

These four arrangements are special in that the resulting MCP will have a value of 0 or be small compared to other types of arrangements (see Figure 5). This is advantageous because after median centering the data, these MCPs will have a negative value and be more natural to color code in the form of a heat map. (A) The MCP of two adjacent ovals whose longest lengths are parallel and whose midpoints are in a line that is perpendicular to the longest lengths will be 0. Three or more consecutive ovals arranged in this pattern will be detected as a straight row of ovals (dotted line box). (B) Similar to A, the MCP of this arrangement of cells will be a small value. Three or more consecutive ovals arranged in this way will be detected as a straight row of ovals (dotted line box). (C & D) The MCP of these two arrangements will be small compared to the other types of arrangements observed in tumors (Figure 5). Three or more consecutive arrangements of these two types will be detected as an arch. The MCP algorithm was specifically formulated to detect rosettes and partial rosettes, both of which exhibit a characteristic arched arrangement of neighboring nuclei (dotted line boxes). (D) Unlike in C, the line connecting the midpoints is not parallel with the horizontal line of this page.

David H. Nguyen. "Detecting Partial Rosettes in Tumor Histopathology Using The Cross Product." (2017) ArXiv [q-bio.TO]

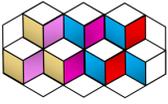
Tissue Spatial Geometrics Laboratory

# Figure 5

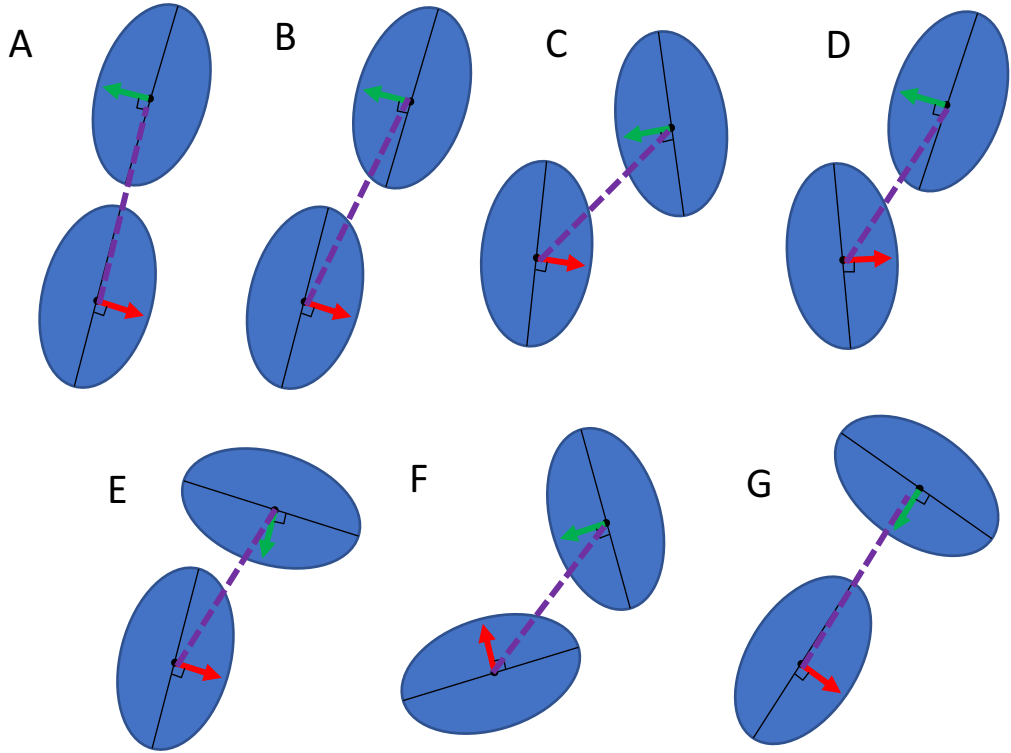

**Figure 5. Neighboring Arrangements That Result in a Large MCP Value.** These arrangements of adjacent nuclei will result in a large MCP value. This is advantageous for quantifying and visualizing rosettes, and partial rosettes, because after subtraction of the median MCP value of a dataset from each individual MCP value (also known as "median centering" the data), the MCP values of these arrangements will still be greater than 0 (positive). However, the MCP values of the special arranges in Figure 4 will be either 0 or less than 0 (negative).

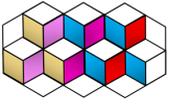

Tissue Spatial Geometrics Laboratory

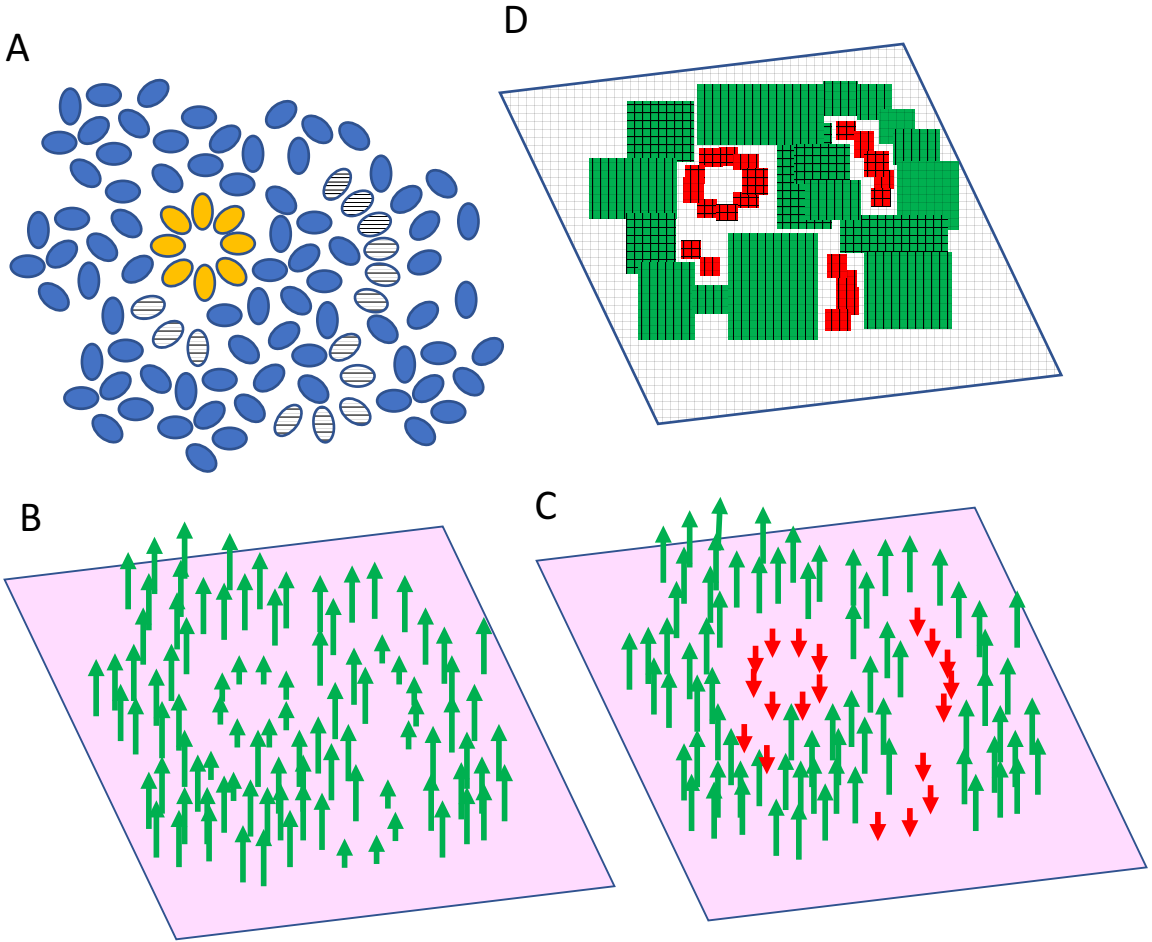

# Figure 6

**Figure 6. The MCP Algorithm Detects Partial Rosettes That Are Difficult for the Human Eye to Perceive.** (A) A cartoon highlighting full rosettes (orange) and partial rosettes (striped) within a collection of non-rosette patterns of neighboring ovals. (B) The cartoon in A is visualized in the form of MCP vectors. The MCP algorithm creates a vector between each neighboring oval. The vector's magnitude is the cross product described in Figure 2. (C) Subtracting the median MCP value of the whole dataset from each individual MCP value in the dataset (i.e. median centering the data), will generate a range of MCP values that straddle 0 (i.e. positive or negative). (D) MCP values in C can be represented as a heat map to help pathologists detect the location of partial rosettes. This method also allows pathologists to stratify images into categories of tumors that have different degrees of partial rosette formation, which is an indication of the degree of differention that is occuring in the tumor but that cannot be assessed by the naked eye.

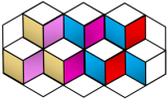
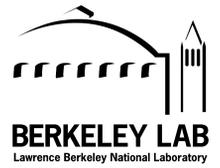